\documentclass[aps,prb,superscriptaddress,reprint]{revtex4-1}

\usepackage{bm}        
\usepackage{amsmath,amssymb,amsthm,upgreek}   
\usepackage{mathtools}
\usepackage{array}
\usepackage{color,soul}
\usepackage{diagbox}
\usepackage{booktabs}
\usepackage{hyperref}
\usepackage{grffile}
\usepackage{graphicx}
\usepackage{float}
\usepackage{chemformula}
\usepackage{gensymb}
\usepackage{multirow}

\newcolumntype{N}{>{\centering\arraybackslash}m{35pt}}
\newcolumntype{G}{>{\centering\arraybackslash}m{155pt}}
\newcolumntype{Y}{>{\centering\arraybackslash}m{133pt}}

\begin{document}
\title{Polarization-resolved Raman spectroscopy of $\alpha$-\ch{RuCl_3} and evidence of room temperature two-dimensional magnetic scattering}

\author{Thuc T. Mai}
\email{thuc.mai@nist.gov}
\affiliation{Department of Physics, The Ohio State University. Columbus, OH 43210, USA}
\affiliation{Nanoscale Device Characterization Division, Physical Measurement Laboratory, NIST, Gaithersburg, MD}
\author{A. McCreary}
\affiliation{Nanoscale Device Characterization Division, Physical Measurement Laboratory, NIST, Gaithersburg, MD}
\author{P. Lampen-Kelley}
\affiliation{Materials Science and Engineering, University of Tennessee, Knoxville, TN}
\affiliation{Materials Science and Technology Division Oak Ridge National Laboratory, Oak Ridge, TN}
\author{N. Butch}
\affiliation{NIST Center for Neutron Research, National Institute of Standards and Technology, Gaithersburg, Maryland 20899, USA}
\affiliation{Center for Nanophysics and Advanced Materials, Department of Physics, University of Maryland, College Park, MD 20742 USA}
\author{J. R. Simpson}
\affiliation{Physics, Astronomy, and Geosciences, Towson University, Towson, MD}
\affiliation{Nanoscale Device Characterization Division, Physical Measurement Laboratory, NIST, Gaithersburg, MD}
\author{J.-Q. Yan}
\affiliation{Materials Science and Technology Division Oak Ridge National Laboratory, Oak Ridge, TN}
\affiliation{Materials Science and Engineering, University of Tennessee, Knoxville, TN}
\author{S. E. Nagler}
\affiliation{Neutron Scattering Division, Oak Ridge National Laboratory, Oak Ridge, TN}
\author{D. Mandrus}
\affiliation{Materials Science and Engineering, University of Tennessee, Knoxville, TN}
\affiliation{Materials Science and Technology Division Oak Ridge National Laboratory, Oak Ridge, TN}
\author{A. R. Hight Walker}
\affiliation{Nanoscale Device Characterization Division, Physical Measurement Laboratory, NIST, Gaithersburg, MD}
\author{R. Vald\'es Aguilar}
\email{valdesaguilar.1@osu.edu}
\affiliation{Department of Physics, The Ohio State University. Columbus, OH 43210, USA}

\date{\today}

\begin{abstract}

Polarization-resolved Raman spectroscopy was performed and analyzed from large, high quality, mono-domain single crystal of $\alpha$-\ch{RuCl_3}, a proximate Kitaev quantum spin liquid. Spectra were collected with laser polarizations parallel and perpendicular to the honeycomb plane. Pairs of nearly degenerate phonons were discovered and show either a 4-fold or 2-fold polarization angle dependence in their Raman intensity, thereby providing evidence to definitively assign the bulk crystal point group as $C_{2h}$.
The low frequency continuum that is often attributed to scattering from pairs of Majorana fermions was also examined and found to disappear when the laser excitation and scattered photon polarizations were perpendicular to the honeycomb plane. This disappearance, along with the behavior of the phonon spectrum in the same polarization configuration, strongly suggests that the scattering continuum is 2-dimensional. We argue that this scattering continuum originates from the Kitaev magnetic interactions that survives up to room temperature, a scale larger than the bare Kitaev exchange energy of approximately 50 K.
\end{abstract}
\maketitle

\section{Introduction}
In the last decade, quantum spin liquids (QSL) research has developed into a burgeoning field in condensed matter physics\cite{Balents2010,KITAEV2006,SavaryBalents2016,HermannsKimchiKnolle2018Review,KnolleMoessner2019,TakagiNagler2019}. Due to the lack of spin ordering and the massively entangled ground state, a QSL is predicted to host a variety of exotic phenomena\cite{SavaryBalents2016}. Because of the exact analytical solution of its ground state and excitations, the Kitaev model\cite{KITAEV2006} is one of the most studied theoretical models of a QSL. Recent experimental and theoretical studies have converged onto a few real systems that could host this model\cite{Banerjee2016,Banerjee2017,Banerjee2018,SandilandsBurch2015,GlamazdaChoi2017,GlamazdaChoi2016,KnollePerkins2014,Perreault2015,Ran2017,ZHOU2018,Rousochatzakis2019}, including $\alpha$-\ch{RuCl_3} and the iridate family of \ch{A_2IrO_3}. In these systems, strong spin-orbit coupling, which leads to $J_{eff}=1/2$, and the honeycomb lattice of the edge-sharing \ch{IrO_6} and \ch{RuCl_6} octahedra give rise to effective bond-dependent exchange interaction of the Kitaev model\cite{Jackeli2009,WinterValenti2016}.

Although long range magnetic order does occur in $\alpha$-\ch{RuCl_3} below 7 K\cite{Sears2017}, inelastic neutron scattering (INS) revealed contributions from both Heisenberg and Kitaev exchange\cite{Banerjee2016}. A magnetic scattering continuum measured by various INS studies\cite{Banerjee2017,Do2017,Banerjee2018} was suggested to arise from excitations of Majorana fermions, a signature of the Kitaev QSL phase\cite{KITAEV2006,KnolleMoessner2019,HermannsKimchiKnolle2018Review}. Others have argued that this continuum comes from magnons that decay rapidly as a consequence of frustrated interactions\cite{winter2017breakdown}. Interestingly, theoretical studies have shown that inelastic scattering with light (such as Raman scattering) can also couple to the fractionalized spin excitations\cite{KnollePerkins2014,Perreault2015,Perreault2016}. The predicted broad Raman scattering signal has been measured in both $\alpha$-\ch{RuCl_3} and $\beta$ and $\gamma$-\ch{Li_2IrO_3}\cite{SandilandsBurch2015,GlamazdaChoi2016,GlamazdaChoi2017}. 

The space group symmetry of bulk crystal $\alpha$-\ch{RuCl_3} is $C 2/m$ (point group $C_{2h}$) at room temperature\cite{CaoNagler2016,JohnsonColdea2015}. The \textbf{a-b} plane consists of Ru$^{3+}$ ions that form a 2-dimensional (2D) honeycomb lattice. The Ru ions reside inside Cl$^-$ octahedra, where due to crystal field splitting and spin orbit coupling, have pseudo-spin values of 1/2\cite{Jackeli2009}. These honeycomb planes are monoclinically stacked along the \textbf{c} axis (Fig. \ref{fig:fig1}(a) and (b)).  The \textbf{c} axis makes an approximately $108\degree$ angle with the \textbf{a} axis while being perpendicular to the \textbf{b} axis, and the van der Waals coupling weakly holds the 2D layers together. The monoclinic stacking breaks the 3-fold rotational symmetry of the honeycomb layers, leaving the 2-fold rotational symmetry around the \textbf{b}-axis as the highest symmetry of the bulk crystal. Refinement of diffraction data from previous studies revealed a small distortion of the honeycomb plane, in the order of $\sim$0.01 \r{A}, effectively also breaking the individual layer's 3-fold symmetry\cite{JohnsonColdea2015,CaoNagler2016}.

\renewcommand{\arraystretch}{1.5}
\begin{table*}[ht]
\centering
\begin{tabular}{N N G G}
                                         &     & \textbf{Parallel} & \textbf{Crossed} \\ \cline{2-4}
\multicolumn{1}{c}{\multirow{2}{*}{\textbf{a-b}}} & $A_g$ & $(a+b) + (a-b)\cos(2\theta)$ & $(b-a)\sin(2\theta)$ \\ \cline{2-4} 
\multicolumn{1}{c}{}                    & $B_g$ &  $e \sin(2\theta)$  &  $e \cos(2\theta)$ \\ \hline
\multicolumn{1}{c}{\multirow{2}{*}{\textbf{a-c}}} & $A_g$ & $(a+c)+(a-c)\cos(2\phi) -2d \sin(2\phi)$ & $(a-c)\sin(2\phi) -2d\cos(2\phi)$ \\ \cline{2-4} 
\multicolumn{1}{c}{}                    & $B_g$ &    N.A.      &   N.A.      \\ \hline
\end{tabular}
\caption{Amplitude of the Raman response of the \textbf{a-b} and \textbf{a-c} plane of $\alpha$-\ch{RuCl_3}, derived as $\mathbf{e}_i \cdot \mathbf{R}_S \cdot \mathbf{e}_s$, where $\mathbf{e}_i$ and $\mathbf{e}_s$ are the incident and scattered polarization directions, respectively, and the raman tensor $\mathbf{R}_S$ are given in Eq. \ref{eq:tensor}. Here the crystalline \textbf{b} axis is the tensor \textbf{y} axis, and \textbf{a} is the \textbf{x} axis, while \textbf{c} is \emph{not} the \textbf{z} axis. N.A. means that no Raman response should occur for that configuration. $\theta$ and $\phi$ are defined in the text.}
\label{tab:RamanResponse}
\end{table*}
\renewcommand{\arraystretch}{1}

Temperature-dependent Raman scattering experiments and theory provided additional evidence that this scattering continuum follows a fermionic behavior at low temperature, as opposed to bosonic\cite{SandilandsBurch2015,GlamazdaChoi2016,Nasu2016,WangBurch2018}. Scattering of a lattice vibrational mode in the proximity of this continuum exhibits a highly asymmetric lineshape\cite{SandilandsBurch2015,GlamazdaChoi2017}. However, a determination of the symmetry and frequency of all of the bulk crystal's Raman phonons has not been uniquely made. Armed with a high quality single crystal sample of $\alpha$-\ch{RuCl_3}, our Raman scattering experiment is able to definitively determine the symmetry of the Raman-active phonons. In addition, multiple phonon peaks, not observed in previous studies, with $\approx$ 2 cm$^{-1}$ separation in frequency were resolved, including the one affected by the continuum. Then, we systematically study the change in the asymmetry of this phonon, with polarization-resolved Raman spectroscopy. We find that the continuum and the phonon asymmetry are not observed when the photon polarizations are perpendicular to the honeycomb plane. This strongly suggests that these effects are of 2D magnetic origin, even at room temperature, a scale larger than the bare Kitaev exchange of $\approx$5--8 meV $\sim$58--90 K \cite{Nasu2016,WangBurch2018,Banerjee2016,WinterValenti2016,yadav2016kitaev}.

\section{Experiment}

\begin{figure}[hb]
\includegraphics[width=.85\columnwidth]{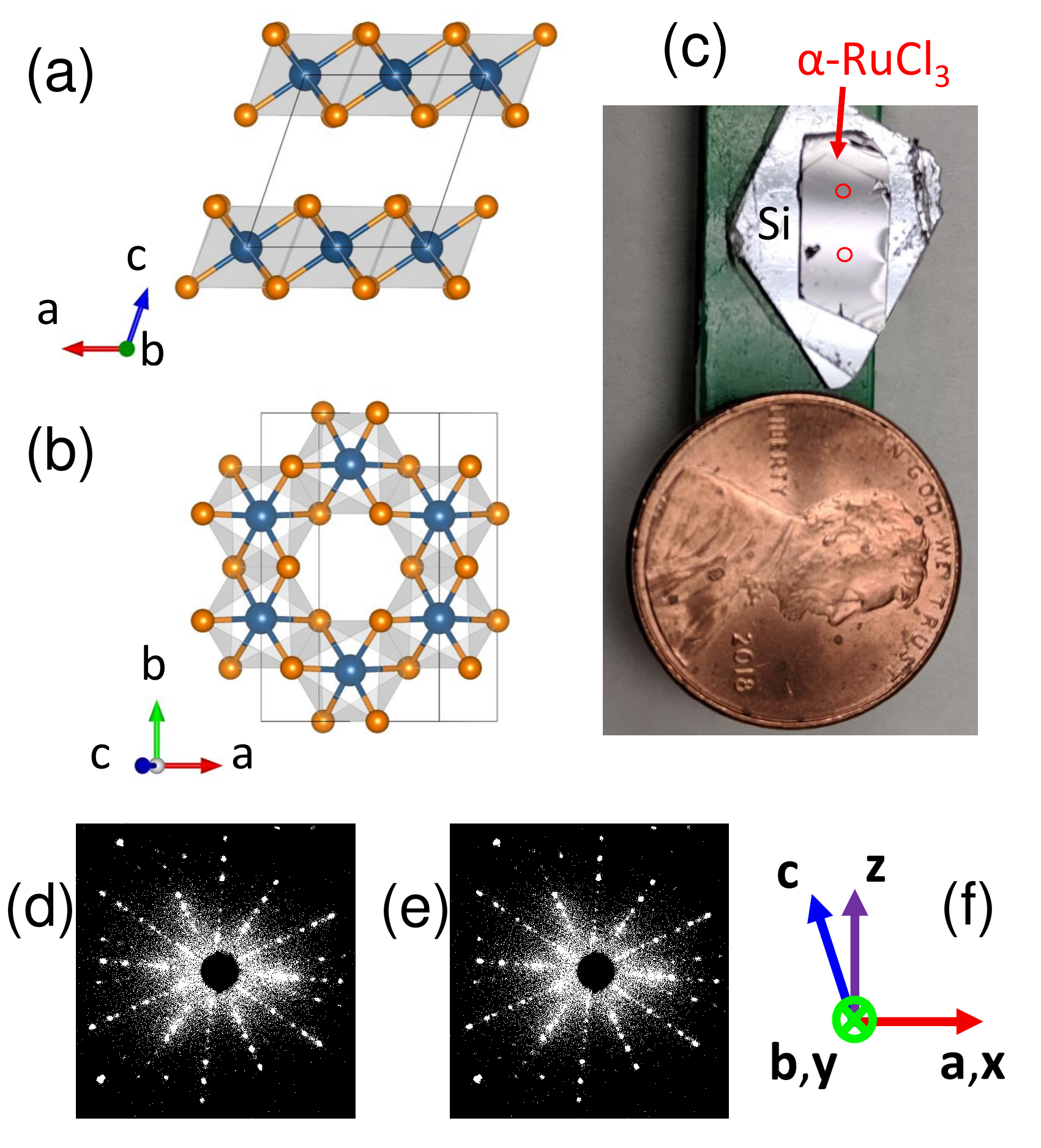}
\caption{(color online) \textbf{Crystal structure and Laue X-ray diffraction pattern}. Crystal structure of $\alpha$-\ch{RuCl_3} projected onto the \textbf{a-c} plane (a) and \textbf{a-b} plane (b). The solid black lines represent the unit cell. The orange balls represent Cl atoms, while the larger blue balls represent Ru atoms. (c) Picture of studied sample; the red circles are the approximate location of where the Laue patterns in (d) and (e) were taken. The well defined circular dots indicate the single-crystallinity of our sample. This pattern remained unchanged over several millimeters on the \textbf{a-b} plane surface. The relationship between the crystallographic axes (a,b,c) and the laboratory (Raman tensor) coordinate (x,y,z) is shown in (f). Notice that \textbf{c} and \textbf{z} are not parallel.}
\label{fig:fig1}
\end{figure}

The single crystal of $\alpha$-\ch{RuCl_3} used here was grown as described elsewhere\cite{CaoNagler2016}. Similar crystals have been studied by neutron scattering\cite{Banerjee2016,Banerjee2018,Banerjee2017}, terahertz\cite{Little2017} and Raman\cite{ZHOU2018,WangBurch2018} spectroscopies. Laue diffraction patterns were taken on various spots across the surface of the sample (\textbf{a-b} plane) at room temperature to confirm the large domain, single crystal nature of our sample (Fig. \ref{fig:fig1}(c), (d), (e)). Raman spectra with 514.5 nm laser excitation were measured in ambient conditions using a single-grating spectrometer (grating with lines density of 1800 mm$^{-1}$, and 1.27 cm CCD detector) in the 180$\degree$ backscattering configuration. A combination of half wave plates and linear polarizers was used to measure spectra in either parallel or crossed polarization configurations, where the polarization orientations were confirmed using \ch{MoS_2} as a reference\cite{Lu2016}. The $\alpha$-\ch{RuCl_3} sample, where the \textbf{a} and \textbf{b} axes were known from Laue diffraction, was placed on a rotating stage to access different crystal orientations. Enabled by the large thickness of our single crystal, we also measured Raman spectra from the \textbf{a-c} plane by mounting the crystal on its side. Integration times ranged between 5 minutes (\textbf{a-b} plane) and 12 minutes (\textbf{a-c} plane), and the laser power was kept below 200 $\rm\upmu W$ through the objective (50x, numerical aperture 0.75 to probe \textbf{a-b} plane, 100x long working distance objective, numerical aperture 0.75 to probe \textbf{a-c} plane) to avoid any local heating of the sample.

Given the point group symmetry of the bulk crystal is $C_{2h}$, we expect 12 Raman active phonons: $6 A_g + 6 B_g$. Group theory predicts the Raman response tensor of $A_g$ and $B_g$ to be the following:

\begin{equation}\label{eq:tensor}
\mathbf{R}_{A_g} = \begin{bmatrix} a & 0 & d \\ 0 & b & 0 \\ d & 0 & c \end{bmatrix} ;
\mathbf{R}_{B_g} = \begin{bmatrix} 0 & e & 0 \\ e & 0 & f \\ 0 & f & 0 \end{bmatrix}
\end{equation}

\noindent where $a, b, c, d, e, f$ are independent coefficients. The $3\times3$ matrices in Eq. \ref{eq:tensor} are written in a coordinate system where the tensor's \textbf{x} and \textbf{y} axes are parallel to the crystallographic \textbf{a} and \textbf{b} axes respectively. The selection rules allow for both $A_g$ and $B_g$ phonons to be measured from the \textbf{a-b} plane. Furthermore, $A_g$ and $B_g$ phonons can be measured simultaneously in \emph{both} crossed and parallel polarization configurations when the polarizations are not coincident with either the \textbf{a} or \textbf{b} crystallographic axis. 

The predicted angular dependence of the Raman response in the backscattering geometry is shown in Table \ref{tab:RamanResponse}, where $\theta$ is the angle between the incoming polarization and the crystal's \textbf{b} axis for the measurement in the \textbf{a-b} plane (inset in Fig. \ref{fig:fig2}), and $\phi$ is the angle away from the \textbf{a} axis for the \textbf{a-c} plane measurement (inset in Fig. \ref{fig:fig5}). The Raman intensity is given by the square of the expressions given in Table \ref{tab:RamanResponse}: $I_{\text{Raman}}=\vert \mathbf{e}_i \cdot \mathbf{R} \cdot \mathbf{e}_s\vert ^2$, where $\mathbf{e}_i$ and $\mathbf{e}_s$ are the incident and scattered polarization directions, respectively. We expect the intensity of the $A_g$ and $B_g$ phonons to be 4-fold modulated, and out of phase with each other as $\theta$ changes within the \textbf{a-b} plane. In contrast in the \textbf{a-c} plane, $B_g$ phonons are not allowed and the $A_g$ phonons should only show 2-fold modulated intensities.

\section{Data}

We obtained multiple Laue X-ray diffraction patterns across our mm sized sample, where two examples are shown in Fig. \ref{fig:fig1} (d) and (e). The consistency of the diffraction patterns at various locations confirmed that our sample is not only of single crystal nature, but also that the domain size is in the order of a few mm. The Laue diffractometer determined the highest symmetry axis of our crystal, the \textbf{b} axis with 2-fold rotational symmetry, in Fig. \ref{fig:fig1}(d), (e). With the \textbf{a-b} plane defined as the basal plane, we can uniquely determine the other two crystal axes, \textbf{a} and \textbf{c}. The unique \textbf{b} axis places constraints on any rank 2 property tensor (i.e. Raman response tensor) by Neumann's principle\cite{birss1964symmetry}. From this constraint, the tensor y-coordinate (in Eq. \ref{eq:tensor}) is identically the crystal's \textbf{b} axis. We then define the \textbf{x} axis to be the \textbf{a} axis. This leaves the laboratory frame's \textbf{z} axis to be different from the crystal's \textbf{c} axis (Fig. \ref{fig:fig1}(f)).

\begin{figure}[b]
\includegraphics[width=1.05\columnwidth]{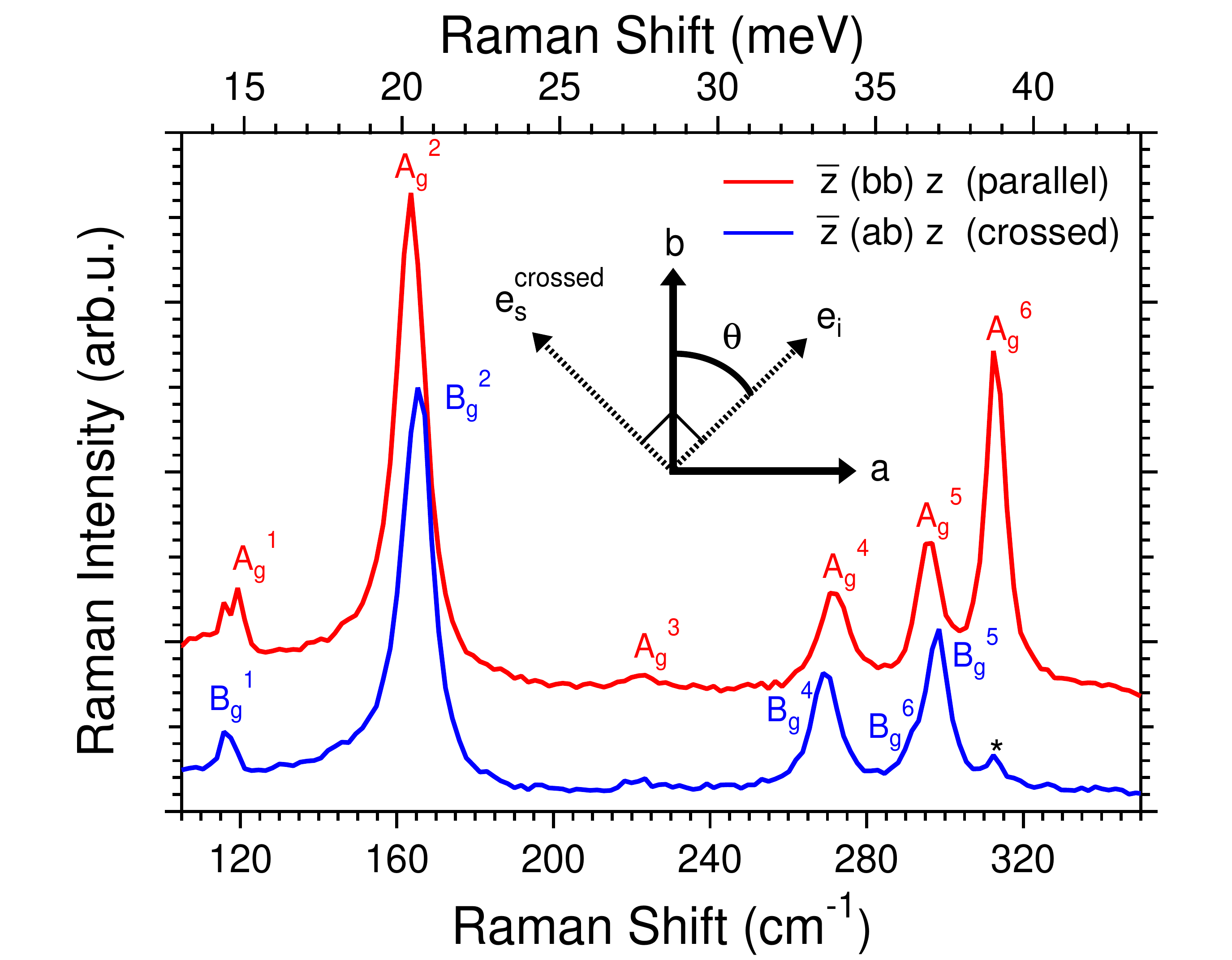}
\caption{\textbf{Raman intensity of the a-b plane}. The parallel configuration data are offset for clarity. Here \textbf{z} is the lab frame direction in which the incident and scattered beams propagate ($\bar{\textbf{z}}= -\textbf{z}$) and \textbf{a} and \textbf{b} are the crystallographic axes. A clear difference is observed between some of the phonon frequencies in the crossed and parallel configuration. We labeled these phonons with $A_g$ and $B_g$, of the bulk point group symmetry $C_{2h}$. The * mode seen in $\bar{\textbf{z}} (\textbf{ab}) \textbf{z}$ is the $A_g^6$ phonon signal leakage due to imperfect polarizers. The inset shows a schematic of the \textbf{a-b} plane measurement, where $\theta$ is defined as the angle between the incident polarization and the \textbf{b} axis.}
\label{fig:fig2}
\end{figure}

Using this knowledge of the crystallographic axes, our initial measurement of the \textbf{a-b} plane collected with incident polarization parallel to \textbf{b} is presented in Fig. \ref{fig:fig2}. The labels $\bar{\textbf{z}} (\textbf{bb}) \textbf{z}$ and $\bar{\textbf{z}} (\textbf{ab}) \textbf{z}$ represent the experimental configurations of the direction of propagation of the incoming beam, the polarization of incident and scattered photons, and the propagation direction of the scattered beam. At first glance, we observed similar spectra to previous Raman studies on $\alpha$-\ch{RuCl_3}\cite{SandilandsBurch2015,GlamazdaChoi2017}. When we carefully compare the two traces in Fig. \ref{fig:fig2}, between the parallel and crossed polarization configuration around 270 cm$^{-1}$ and 295 cm$^{-1}$, there are clear differences in the phonon peak frequencies of approximately 2 cm$^{-1}$. Similarly, two distinct and weaker phonons can be seen around 119 cm$^{-1}$. Our measurements show what appears to be two asymmetric phonon peaks around 164 cm$^{-1}$ (Fig. \ref{fig:fig2}) and a broad background. An asymmetric Fano lineshape typically arises when a discrete resonance, a phonon in this case, is coupled to a broad continuum\cite{Fano1961}. This scattering continuum in $\alpha$-\ch{RuCl_3} has been subjected to multiple low temperature studies\cite{SandilandsBurch2015,GlamazdaChoi2017}, and its origin at room temperature will be discussed in the next section.

To determine the phonon frequencies, the peaks in the Raman spectra were fit with Lorentzian lineshapes, with the exception of the asymmetric peaks near 164 cm$^{-1}$, where the Fano lineshape:
\begin{equation}\label{eq:Fano}
I (\omega) =  I_0 \frac{1}{(1+q^2)}\frac{(q+(\frac{\omega - \omega_0}{\Gamma}))^2}{1+(\frac{\omega - \omega_0}{\Gamma})^2}
\end{equation}
\noindent was used instead. Here, $q$ represents the asymmetry parameter. A Lorentzian lineshape is recovered as $q$ goes to infinity (or $1/|q| = 0$). $I_0$ is an overall intensity scale, $\Gamma$ is the intrinsic width (inverse lifetime) of the phonon, and $\omega_0$ is the natural frequency of the mode. We label the phonons with the irreducible representations $A_g$ and $B_g$ of the point group $C_{2h}$ in Tab. \ref{tab:RamanModes} (and Fig. \ref{fig:fig2}) in contrast to previous reports. The Fano parameters ($1/|q|$) found for parallel ($A_g^2$) and crossed ($B_g^2$) configuration are $0.064$ and $0.091$, respectively.

\renewcommand{\arraystretch}{1.5}
\begin{table*}[ht]
\centering
\begin{tabular}{|Y|N|N|N|N|N|N|}
\hline
\diagbox{Symmetry}{Phonon number}      &  1$^\dagger$    &   2$^\dagger$   &  3  &   4$^\dagger$   &   5$^\dagger$   &   6   \\ \hline
$A_g$ & 119.4 & 164.2 & 222.0 & 271.9 & 295.8 & 313.1 \\ \hline
$B_g$ & 115.8 & 165.8 & N.O.  & 269.3 & 297.6 & 291.5 \\ \hline
\end{tabular}
\caption{List of $A_g$ and $B_g$ phonon frequencies in units of cm$^{-1}$. The symmetries of the phonons were determined by their intensities when $\textbf{e}_i$ or $\textbf{e}_s$ were along a crystalline axis. In such polarization configuration, $A_g$ modes appear in parallel and $B_g$ in crossed. The $^\dagger$ denotes pairs of nearly degenerate phonon modes, not observed in previous publications. $B_g^6$ was deliberately labeled out of order to make the nearly degenerate phonon pairs clearer. N.O. means not observed.}
\label{tab:RamanModes}
\end{table*}
\renewcommand{\arraystretch}{1}

\begin{figure}[b]
\includegraphics[width=1\columnwidth]{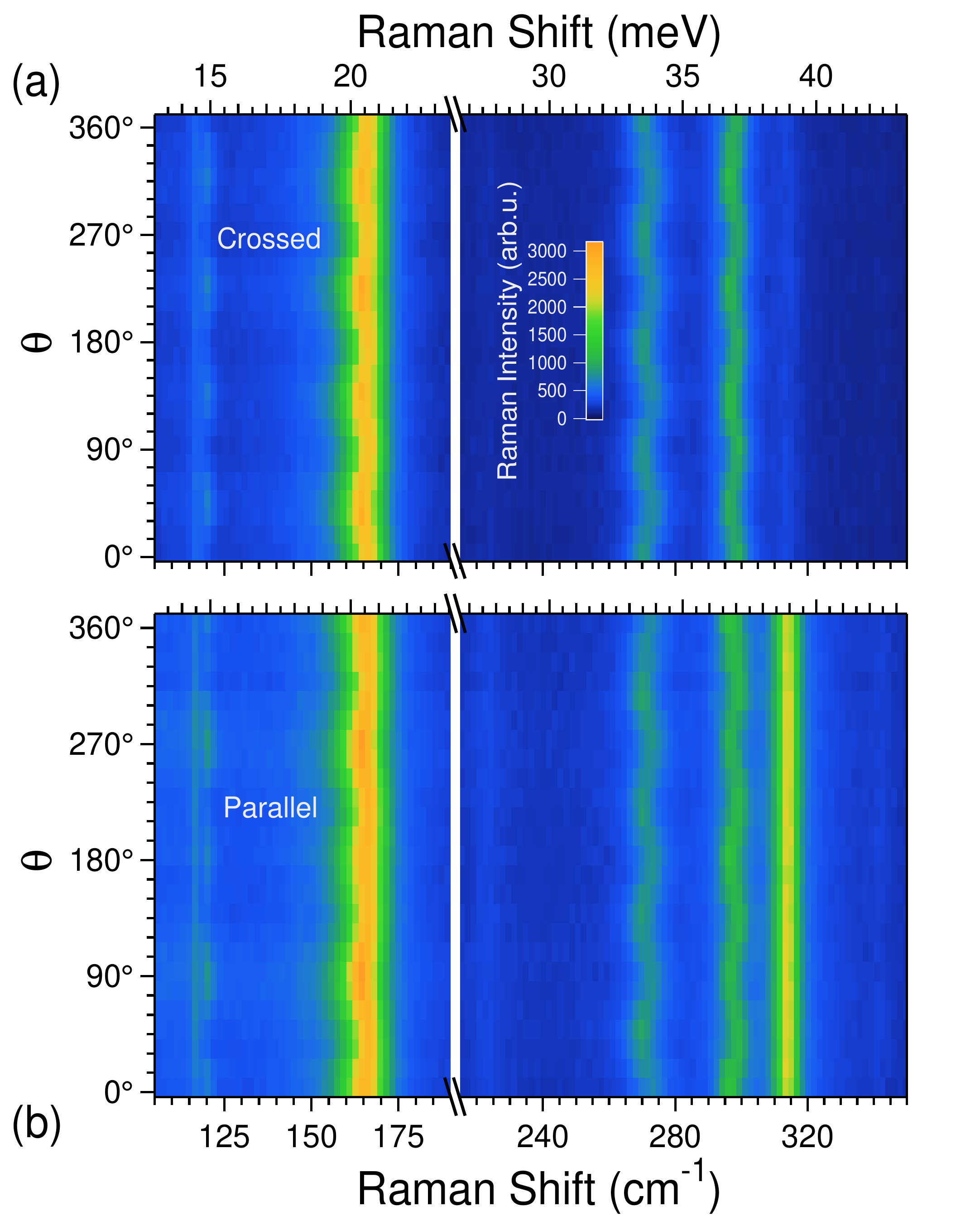}
\caption{\textbf{Raman spectra in the a-b plane as a function of rotation angle} ${\theta}$. False color maps of the Raman intensity as a function of frequency and angle $\theta$, shown for the crossed (a), and parallel (b), configurations. $\theta = 0$ corresponds to $\mathbf{e}_i ,\mathbf{e}_s \parallel \mathbf{b}$ in the parallel configuration, and $\mathbf{e}_i \parallel \mathbf{a}$ and $\mathbf{e}_s \parallel \mathbf{b}$ in the crossed configuration.}
\label{fig:fig3}
\end{figure}

\begin{figure*}[t]
\includegraphics[width=0.7\textwidth]{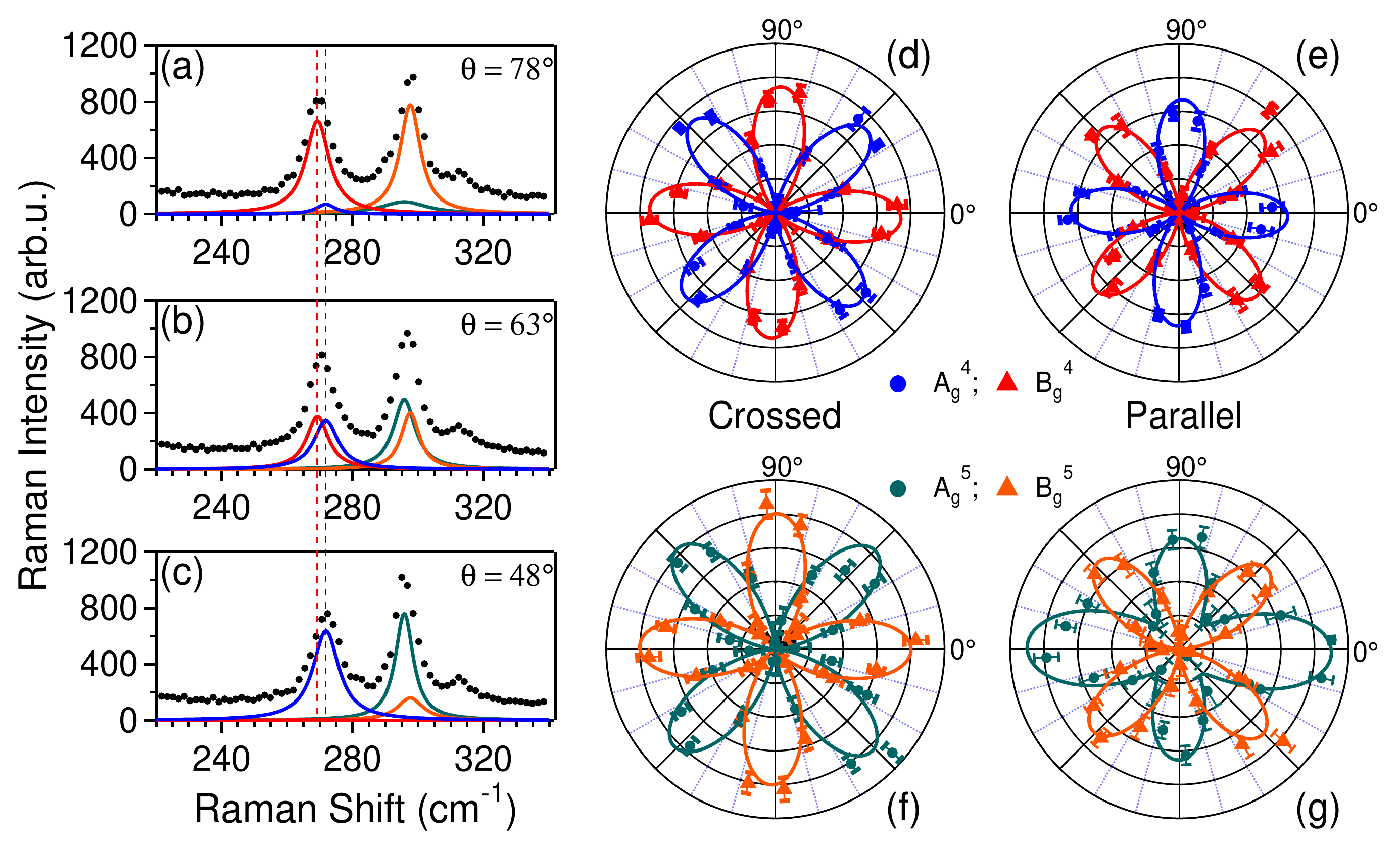}
\caption{\textbf{Angular dependence of $A_g^4$, $B_g^4$, $A_g^5$, $B_g^5$ phonons}. Panels (a), (b), (c) show the raw data (dots) and fitted Lorentzian (lines) at certain angles in the crossed polarization configuration. The spectral weight outputs from the data fit are shown in (d) and (e) ((f) and (g)) for $A_g^4$ and $B_g^4$ ($A_g^5$ and $B_g^5$) for crossed ((d) and (f)) and parallel ((e) and (g)) polarization configurations. The solid lines in ((d)-(g)) are obtained by fitting the spectral weights with the modulus squared of the corresponding expressions found in Table \ref{tab:RamanResponse}. All polar plots are on the same radial scale. The error bars reprensent one standard deviation, obtained by the fitting procedure. }
\label{fig:fig4} 
\end{figure*}

With fixed polarizers and rotating the sample, we collected Raman spectra as a function of the angle between the laser polarizations and the crystalline axes. In the \textbf{a-b} plane, oscillations in the scattered intensity in both crossed and parallel configuration (Fig. \ref{fig:fig3}), around 164 cm$^{-1}$, 270 cm$^{-1}$, and 295 cm$^{-1}$ were observed. In contrast, the $A_g^6$ phonon remained constant in frequency. Each peak frequency oscillation can be explained by a pair of A$_g$ and B$_g$ phonons. The phonons in each pair change in intensity out of phase with each other as the sample is rotated which results in the apparent oscillation of a phonon peak frequency. We employed a fitting model that contains Lorentzians with frequencies fixed by the $A_g$ and $B_g$ phonon frequencies. Three example spectra (from crossed polarizations configuration) and their fitted curves are plotted in Figure \ref{fig:fig4}. The spectral weight of the $A_g^4$, $B_g^4$, $A_g^5$, $B_g^5$ phonons are also plotted as a function of the angle (Fig. \ref{fig:fig4}). The solid lines are group theory predicted responses, Table \ref{tab:RamanResponse}, which are in excellent agreement with our data. In order to measure only the $A_g$ or $B_g$ phonons in the \textbf{a-b} plane, the incident and scattered polarizations have to be parallel to either the \textbf{a} or \textbf{b} axis of the crystal.

Our angular dependent measurements of the \textbf{a-c} plane confirmed the group theoretical prediction, showing no oscillating phonon peaks. In the backscattering geometry, only $A_g$ phonons are active in this plane (see eq. \ref{eq:tensor}). Surprisingly, the $A_g^2$ phonon showed a dramatic shift in the frequency of its maximum intensity as a function of sample orientation $\phi$ (Fig. \ref{fig:fig5}). Additionally, its asymmetry changed dramatically, becoming nearly symmetric at certain angles, e.g. $1/|q| \rightarrow 0$ when $\phi = 90\degree$ (see inset of Fig. \ref{fig:fig5}). This angle corresponds to the incident and scattered laser polarizations being perpendicular to the \textbf{a-b} plane. Also at this angle, the background/scattering continuum below the phonon frequency vanishes.

\section{Discussion}

\subsection{The effects of symmetry reduction on Raman phonons}

The observations above (the frequency separation of the nearly degenerate phonons, and the polarization-dependent details that allow us to determine the point group of the crystal), are made possible only by the large, high quality, single crystal sample with crystallographic domain size larger than a few mm. Previous studies\cite{SandilandsBurch2015,GlamazdaChoi2017,Mashhadi2018,ZHOU2018,WangBurch2018,Li2019} did not report multiple phonons with $\approx 2$ cm$^{-1}$ difference in their frequencies between crossed and parallel configuration. In these studies, the authors assumed the point group $D_{3d}$ of a single perfect honeycomb layer to label the $\alpha$-\ch{RuCl_3} Raman phonon spectra with $4E_g + 2A_{1g}$. We postulate that the samples used in those studies either contain multiple domains in the volume probed by their laser, or the laser polarizations are not parallel to a crystalline axis. In either case, broader, averaged phonon peaks would appear in place of the separate $A_g$ and $B_g$ modes as seen in Fig. \ref{fig:fig4} (b). It is interesting to see how the conclusions from these previous experiments would change with the full details of the Raman phonon spectrum. 

By comparing our result to other Raman studies\cite{SandilandsBurch2015,Mashhadi2018,ZHOU2018,Li2019}, we noticed that the pairs of phonons $A_g^1$ and $B_g^1$, $A_g^2$ and $B_g^2$, $A_g^4$ and $B_g^4$, $A_g^5$ and $B_g^5$ were labeled as $E_g$ modes. Factor group analysis reveals that the doubly degenerate $E_g$ phonons of $D_{3d}$, the often assumed symmetry of $\alpha$-\ch{RuCl_3} crystals, would split into the $A_g$ and $B_g$ modes of $C_{2h}$, the point group symmetry of bulk $\alpha$-\ch{RuCl_3}\cite{JohnsonColdea2015,CaoNagler2016}. Our observation of the nearly degenerate $A_g$ and $B_g$ phonons confirms this picture. It also suggests that the energy scale of the symmetry breaking effect that splits these modes is small $\lesssim$1 meV.

Similar splitting of the $E_g$ modes in the phonon spectra were calculated by density functional theory\cite{LarsonKaxiras2018} and confirmed experimentally\cite{Djurdji2018} in CrI$_3$, a material that shares the same point group symmetry with $\alpha-$RuCl$_3$ at room temperature. Larson and Kaxiras\cite{LarsonKaxiras2018} conclude that the energy differences between the $A_g$ and $B_g$ pairs of phonons come from the weak interlayer coupling. This is consistnet with the small frequency differences that we have measured. We also expect the distortion of each honeycomb layer, where two of the bonds of the hexagon differ from the other four, to play a similar role in the phonon splitting. Although the magnitude of this lattice distortion is minute ($\approx0.2\%$)\cite{CaoNagler2016,JohnsonColdea2015}, it is not inconsistent with our results. Further studies are necessary to clarify the contribution from each mechanism responsible for the observed splitting.

\subsection{Scattering continuum and the Fano lineshape}

\begin{figure}[h]
\includegraphics[width=0.9\columnwidth]{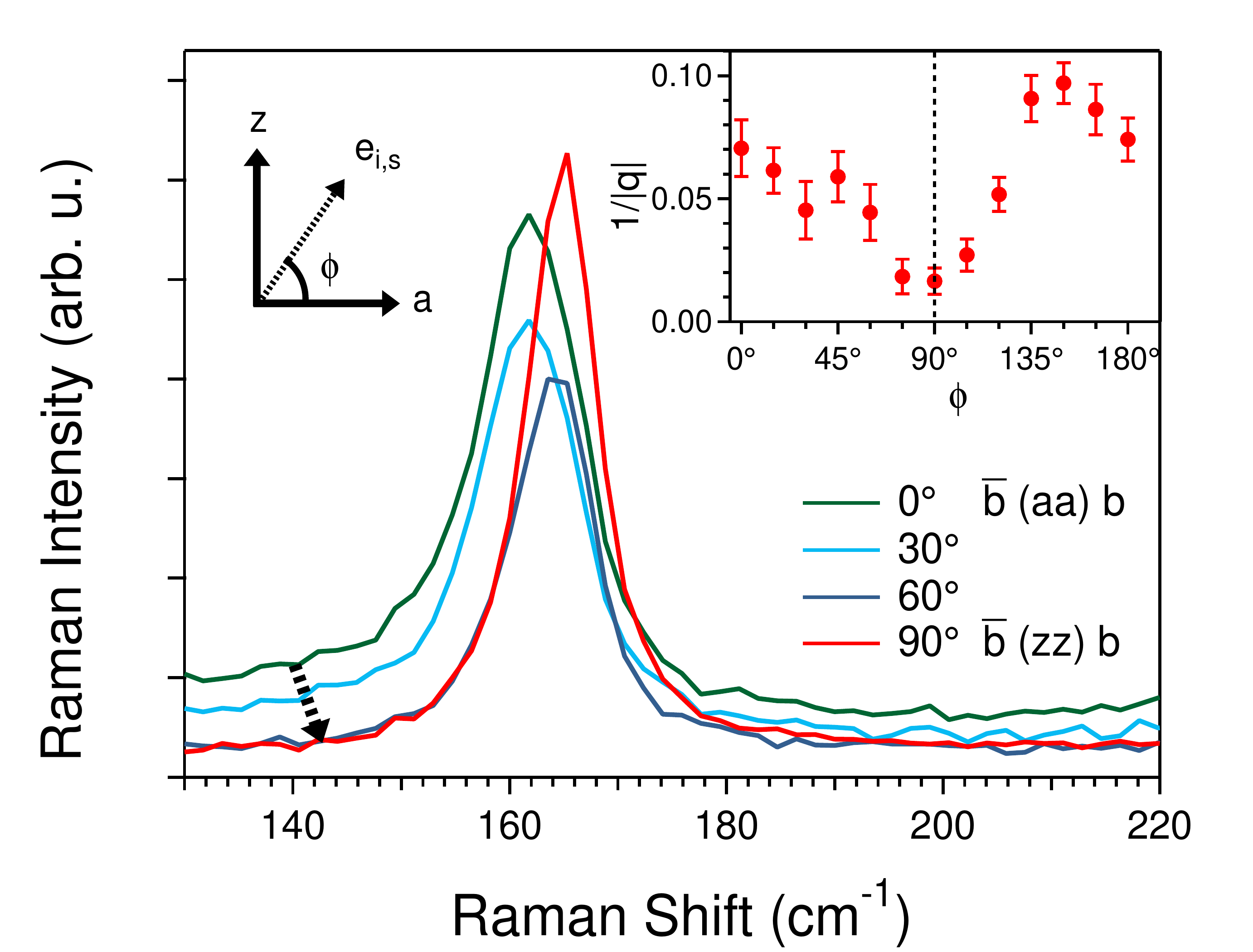}
\caption{\textbf{Changes in the asymmetry of the $A_g^2$ phonon}. Here, $\phi$ is defined as the angle the incoming polarization makes with the \textbf{a} axis, as shown in the left inset. When the laser polarizations are perpendicular to the \textbf{a-b} plane ($\phi=$90$\degree$), the $A_g^2$ phonon appears to be fully symmetric. The dashed arrow highlights both the change in asymmetric lineshape and the disappearance of the scattering continuum. The phonon frequencies obtained by fitting $\phi =0\degree$ and $90\degree$ differs by about $2.5$ cm$^{-1}$. The inset shows the Fano parameter $1/q$. We expect $1/q$ to vanish when the phonon becomes fully symmetric.}
\label{fig:fig5} 
\end{figure}

Previous reports have characterized the Raman intensity around 164 cm$^{-1}$ with a Fano peak and a broad scattering continuum. Our measurements of the \textbf{a-b} plane show that there are two Raman phonons near 164 cm$^{-1}$, with two different asymmetry parameters $1/|q|$ as shown in Figure \ref{fig:fig3}. It was argued that at low temperature, such scattering continuum comes from the magnetism in $\alpha$-\ch{RuCl_3}\cite{SandilandsBurch2015,GlamazdaChoi2017}. In this picture, it is expected that the Raman scattering process measures the excitation of the itinerant Majorana fermions in the Kitaev QSL, and can appear as a broad feature in the spectrum\cite{KnollePerkins2014,Perreault2015,Perreault2016,Nasu2016,HermannsKimchiKnolle2018Review,Rousochatzakis2019}. Our observation in the \textbf{a-c} plane (Fig. \ref{fig:fig5}) where only the $A_g^2$ phonon is allowed, showed three dramatic changes in the phonon Raman response as the sample was rotated around the \textbf{b} axis: the shift of the phonon peak frequency, the vanishing of the continuum and of $1/|q|$ when the laser polarization is aligned perpendicular to the honeycomb plane. The Fano line shape is expected to show a large shift in the measured peak frequency when the energy of the discrete state is comparable to the energy of the background\cite{Fano1961}. Indeed, we observed a difference of about $2.5$ cm$^{-1}$ in the bare phonon frequency of $A_g^2$ from fitting with eq. \ref{eq:Fano} at $\phi = 0\degree$ and $90\degree$. Therefore, the Fano line shape and the continuum share the same origin, as they disappear together when $\phi =90\degree$. These phenomena are consistent with the idea that the continuum cannot be excited with $\mathbf{e}_{i,s} \perp$ \textbf{a-b} plane, and thus suggests that the scattering continuum seen in our data is a 2D effect.

While the continuum disappears, the $A_g$ phonons are still present. Combined with the presence of the continuum when $B_g$ phonons are not active ($\mathbf{\bar{b} (aa) b}$), we conclude that the continuum cannot come from any phononic contribution of the $A_g$ or $B_g$ modes. We can also rule out the single-particle excitation of free electrons contributing to the continuum. We would expect the scattering of electrons between the weakly coupled van der Waals layers to be weak, leading to a 2D-like quasi-elastic response. However, this material is a Mott insulator, with an optical gap of about 1 eV\cite{Plumb2014,Sinn2016,Sandilands2016,Reschke2018}. Temperature dependent electrical resistivity measurement further confirm the insulating behavior of $\alpha$-\ch{RuCl_3}\cite{Mashhadi2018}. The lack of free carriers is also evident in various far-infrared and terahertz transmission spectroscopy experiments, especially through $\sim$ mm thick single crystals of $\alpha$-\ch{RuCl_3}\cite{Little2017,WangLoidl2017,Reschke2017,Reschke2018}. In addition, another candidate to host the Kitaev QSL phase, $\beta$-\ch{Li_2IrO_3}, showed similar broad scattering background at low frequency along with multiple asymmetric phonons, even at room temperature\cite{GlamazdaChoi2016}. 

There have been several studies of Raman scattering in the Kitaev model. They predict a scattering continuum coming from the excitation of the Majorana fermions, in both the gapped, and gapless Kitaev QSL ground states\cite{KnollePerkins2014,Perreault2015,Nasu2016,Perreault2016,HermannsKimchiKnolle2018Review,Rousochatzakis2019}. In particular, \citet{Perreault2016} showed that Raman scattering from Kitaev exchange is only absent for incoming and outgoing polarizations perpendicular to the honeycomb plane. At the same time, the response in the $\mathbf{a}-\mathbf{c}$ plane is not zero if both the 3-fold symmetry of the perfect honeycomb lattice is broken, and both direct and mediated superexchange are considered in the Kitaev interaction. Our observations support this picture, as the phonon asymmetry and the continuum are present in all polarization combinations, except when they both point perpendicular to the $\mathbf{a}-\mathbf{b}$ plane, and by the fact that our $A_g-B_g$ phonon splitting means that the 3-fold symmetry is broken. Furthermore, it is expected that the Kitaev scattering continuum persist up to temperatures of at least $7\times J_K$, where $J_K$ is the isotropic Kitaev exchange constant\cite{Nasu2016,WangBurch2018}. The value of $J_K$ has been estimated to be 5 meV to 8 meV (58 K to 93 K)\cite{SandilandsBurch2015,Banerjee2016,Banerjee2016,WinterValenti2016,yadav2016kitaev}, which places the continuum persistence temperature well above 300 K. Combined with the polarization-resolved results described above, it strongly suggests that the scattering continuum measured in bulk $\alpha$-\ch{RuCl_3} comes from the Kitaev QSL physics at room temperature.

\section{Summary}
Using polarization-sensitive Raman scattering of the phonon spectrum in $\alpha$-\ch{RuCl_3}, we showed that the point group symmetry of the crystal is $C_{2h}$, and not the $D_{3d}$ point group of its perfect honeycomb layer. Because of the large, high-quality single crystal used in this experiment, we correctly identified and labeled the full Raman active phonon spectrum. We observed a broad scattering continuum and two Fano line shape resonances near 164 cm$^{-1}$. The scattering continuum and the Fano asymmetry were found to vanish with laser polarizations perpendicular to the honeycomb plane. We argued that the Kitaev exchange is the most likely cause for the continuum and the asymmetry even at room temperature. Thus, we provide an example that shows the power of polarization-resolved Raman scattering in the study of Kitaev materials.

\section*{Acknowledgements}
Funding for this work was provided by OSU's Center for Emergent Materials, an NSF MRSEC, under award number DMR-1420451. T.T.M., A.M., and A.R.H.W. would like to acknowledge the National Institute of Standards and Technology (NIST)/National Research Council Postdoctoral Research Associateship Program and NIST-STRS (Scientific and Technical Research Services) for funding. P.L.K. and D.M. acknowledge support from the Gordon and Betty Moore Foundation’s EPiQS Initiative through grant GBMF441. J.Y. was supported by the DOE Office Basic Energy Sciences, Materials Sciences and Engineering Division. S.E.N. was supported by the US DOE, Division of Scientific User Facilities, Office of Basic Energy Sciences. We acknowledge useful discussions with N.P. Armitage, Y-M. Lu and N. Trivedi.

\bibliography{alphaRuCl3_final}
\end{document}